\documentclass[twocolumn,aps,prd,preprintnumbers,bibnotes11pt,superscriptaddress,amsmath,amssymb,nofootinbib,floatfix]{revtex4-2}

\usepackage{graphicx,subfigure}
\usepackage{float}
\usepackage{color}
\usepackage{epstopdf}
\usepackage{amsmath}
\usepackage{amsthm}
\usepackage{amssymb}
\usepackage{amsfonts}
\usepackage{graphicx}
\usepackage{txfonts}
\usepackage{ulem}
\usepackage{mathtools}
\usepackage{bm}
\usepackage{array,multirow}

\usepackage{dcolumn}

\usepackage[colorlinks=true,urlcolor=blue,citecolor=blue,linkcolor=red]{hyperref}
\usepackage{booktabs}
\usepackage{braket}
\usepackage[dvipsnames]{xcolor}
\usepackage{ulem}

\usepackage{amsmath}
\usepackage{graphicx}
\usepackage{ulem}
\usepackage{dcolumn}
\usepackage{epsfig}
\usepackage{bm}
\usepackage{array}
\usepackage{hyperref}
\hypersetup{
colorlinks=true,
citecolor=blue,
linkcolor=blue,
urlcolor=black,
pdfmenubar=true
}

\usepackage{hyperref}
\usepackage{graphicx}
\usepackage{amsmath}
\usepackage{amsfonts}
\usepackage{amssymb}
\usepackage{xcolor}
\usepackage{bbm}
\usepackage{calrsfs}
\usepackage{dutchcal}
\usepackage{amsthm}
\usepackage{tikz}
\usetikzlibrary {arrows.meta}

\newcommand{\new}[1]{\textcolor{blue}{#1}}

\newcommand{\be}{\begin{equation}}
\newcommand{\ee}{\end{equation}}

\newcommand{\vect}[1]{\boldsymbol{#1}}
\newcommand{\makeref}[1]{(\ref{#1})}

\newcommand{\np}{\mathcal{N}}

\newcommand{\NT}{N}

\allowdisplaybreaks

\begin{document}

\title{Squeezing-enhanced accurate differential sensing under large phase noise}

\author{Robin Corgier$^\dagger$}
\affiliation{LTE, Observatoire de Paris, Université PSL, Sorbonne Université, Université Lille, LNE, CNRS, 61 avenue de l’Observatoire, 75014 Paris, France}

\author{Marco Malitesta$^\dagger$}
\affiliation{Istituto Nazionale di Ottica, Consiglio Nazionale delle Ricerche (INO-CNR), Largo Enrico Fermi 6, 50125 Firenze, Italy.}
\affiliation{Department of Physics and Astronomy, Universita di Firenze, Via Sansone 1 50019 Sesto Fiorentino, Italy}

\author{Leonid A. Sidorenkov}
\affiliation{LTE, Observatoire de Paris, Université PSL, Sorbonne Université, Université Lille, LNE, CNRS, 61 avenue de l’Observatoire, 75014 Paris, France}

\author{Franck Pereira Dos Santos}
\affiliation{LTE, Observatoire de Paris, Université PSL, Sorbonne Université, Université Lille, LNE, CNRS, 61 avenue de l’Observatoire, 75014 Paris, France}

\author{Gabriele Rosi}
\affiliation{Department of Physics and Astronomy, Universita di Firenze, Via Sansone 1 50019 Sesto Fiorentino, Italy}
\affiliation{Istituto Nazionale di Fisica Nucleare (INFN), Via Sansone 1, 50019 Sesto Fiorentino, Italy}

\author{Guglielmo M. Tino}
\affiliation{Istituto Nazionale di Ottica, Consiglio Nazionale delle Ricerche (INO-CNR), Largo Enrico Fermi 6, 50125 Firenze, Italy.}
\affiliation{Department of Physics and Astronomy, Universita di Firenze, Via Sansone 1 50019 Sesto Fiorentino, Italy}
\affiliation{Istituto Nazionale di Fisica Nucleare (INFN), Via Sansone 1, 50019 Sesto Fiorentino, Italy}
\affiliation{European Laboratory for Nonlinear Spectroscopy (LENS), Via Nello Carrara 1, 50019 Sesto Fiorentino, Italy}

\author{Augusto Smerzi}
\affiliation{Istituto Nazionale di Ottica, Consiglio Nazionale delle Ricerche (INO-CNR), Largo Enrico Fermi 6, 50125 Firenze, Italy.}
\affiliation{European Laboratory for Nonlinear Spectroscopy (LENS), Via Nello Carrara 1, 50019 Sesto Fiorentino, Italy}
\affiliation{QSTAR, Largo Enrico Fermi 2, 50125 Firenze, Italy.}

\author{Leonardo Salvi$^*$}
\affiliation{Department of Physics and Astronomy, Universita di Firenze, Via Sansone 1 50019 Sesto Fiorentino, Italy}
\affiliation{Istituto Nazionale di Fisica Nucleare (INFN), Via Sansone 1, 50019 Sesto Fiorentino, Italy}
\affiliation{European Laboratory for Nonlinear Spectroscopy (LENS), Via Nello Carrara 1, 50019 Sesto Fiorentino, Italy}

\author{Luca Pezz$\grave{\text{e}}^*$}
\affiliation{Istituto Nazionale di Ottica, Consiglio Nazionale delle Ricerche (INO-CNR), Largo Enrico Fermi 6, 50125 Firenze, Italy.}
\affiliation{European Laboratory for Nonlinear Spectroscopy (LENS), Via Nello Carrara 1, 50019 Sesto Fiorentino, Italy}
\affiliation{QSTAR, Largo Enrico Fermi 2, 50125 Firenze, Italy.}

\begin{abstract}

Atom interferometers are reaching sensitivities fundamentally constrained by quantum fluctuations. 
A main challenge is to integrate entanglement into quantum sensing protocols to enhance precision while ensuring robustness against noise and systematics.
Here, we investigate differential phase measurements with two atom interferometers using spin-squeezed states, accounting for common-mode phase noise spanning the full $2\pi$ range.
We estimate the differential signal using model-free ellipse fitting, a robust method requiring no device calibration and resilient to additional noise sources.
Our results show that spin-squeezing enables sensitivities below the standard quantum limit.
Specifically, we identify optimal squeezed states that minimize the differential phase variance, scaling as $N^{-2/3}$, while eliminating bias inherent in ellipse fitting methods.
We benchmark our protocol against the Cramér-Rao bound and compare it with hybrid methods that incorporate auxiliary classical sensors.
Our findings provide a pathway to robust and high-precision atom interferometry, in realistic noisy environments and using readily available states and estimation methods.

\end{abstract}

\maketitle
\def\thefootnote{$\dagger$}\footnotetext{These authors contributed equally to this work.}
\def\thefootnote{$*$}\footnotetext{Corresponding authors leonardo.salvi@unifi.it and luca.pezze@ino.cnr.it.}

\section{Introduction}
 
Precision measurements with atomic devices are often achieved through interferometry techniques, where a physical quantity of interest (e.g. an acceleration, a force, or a frequency) is encoded into a detectable phase shift~\cite{CroninRMP2009, VarennaProceedings, SafronovaRMP2018, Bongs2019, Geiger20}.
Atom interferometers are now reaching sensitivities fundamentally bounded by quantum fluctuations, in particular given by the standard quantum limit (SQL)~\cite{Gauguet09, Sorrentino2014, Janvier22}, which applies when uncorrelated atoms are used~\cite{Pezze2009}.
To surpass the SQL, several proof-of-principle experiments have explored the engineering of quantum correlations among atoms~\cite{Pezze18}. 
Such correlations are often generated through a spin-squeezing process \cite{Pezze18, Ma2011, Kitagawa1993, Wineland1994}. 
This consists in reducing fluctuations in a collective pseudo-spin observable, leading to smaller measurement uncertainty and thus higher precision in sensing.
So far, the creation of spin-squeezed states has been reported in atomic clocks~\cite{Louchet-ChauvetNJP2014, Kruse16, Hosten16, Laudat18, Pedrozo20, Huang23}.
In particular, an optical clock frequency measurement below the SQL has been realized~\cite{Robinson22}.
Entangled sources have also been proposed for inertial sensors~\cite{Salvi18, Szigeti20, Corgier21a, Corgier21b, Corgier23}.
Delocalization of entanglement in a superposition of well-separated atomic momentum states has been demonstrated with ultracold atoms~\cite{Graham22,Malia22} and Bose-Einstein condensates~\cite{Anders21}, leading to the realization of an atomic gravimeter with sensitivity below the SQL~\cite{Klempt24}.
However, these pioneering results were achieved in controlled environments with low phase noise. 
In practice, stochastic phase fluctuations can displace squeezed states from their optimal operating point: this causes the antisqueezed quadrature to leak in the measurement and deplete  the advantage gained over coherent states~\cite{AndrePRL2004, Braverman2018}.
Current efforts focus on mitigating this issue, including optimized feedback loops with multiple squeezed ensembles~\cite{PezzePRL2020, PezzePRXQQ2021}, non-demolition measurements \cite{BorregaardPRL2013} or variational optimizations of the probe state ~\cite{KaubrueggerPRX2021, MarciniakNATURE2021}.

\begin{figure*}[t!]
\includegraphics[width=\textwidth]{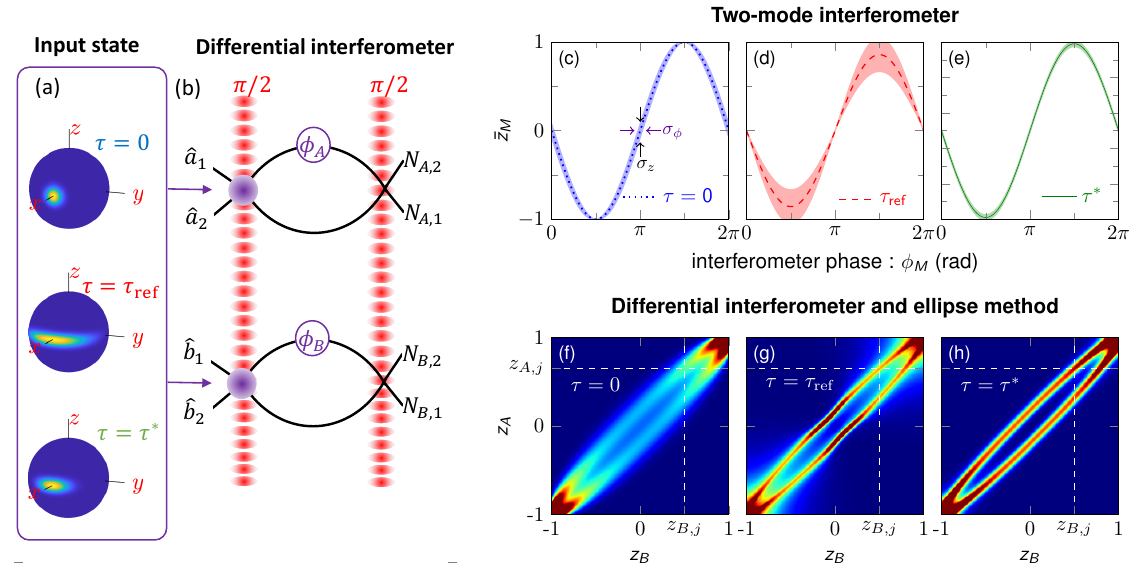}
\caption{Differential interferometer scheme.
(a) Wigner distributions of the two-mode input states of each interferometer considered in this manuscript: 
(top) a coherent state given by Eq.~(\ref{Eq_psi_Coh}) and corresponding to $\tau=0$;
(middle) a spin-squeezed state given by Eq.~(\ref{Eq_psi_Sq}) with strength $\tau_{\rm ref}$; 
(bottom) a spin-squeezed state with strength $\tau^*$. %
(b) The differential scheme consists of two interferometers interrogated simultaneously by a common laser. 
The accumulated phase differences $\phi_A$ and $\phi_B$ are read-out by detecting the particle numbers at the output ports of each interferometer.
(c-e) Mean value of the normalized population imbalance, $\bar{z}_M$, as a function of the single interferometer phase shift $\phi_M$, for $\tau = \{0, \tau_{\rm ref}, \tau^*\}$. 
The shaded area is the corresponding mean square fluctuation. 
The different panels correspond to the different input states in (a).
(f-h) Probability distribution Eq.~(\ref{joint_probability}) for a phase noise spanning the full $\phi_{\rm cn}\in[0, 2\pi]$ interval, and different probe states: (f) two coherent states, (g-h) two spin-squeezed states with strength $\tau_{\rm ref}$ (g) and $\tau^*$ (h). 
Here, $\delta\phi=\pi/16$ and  
$\NT=100$.
}
\label{fig_1}
\end{figure*}

This work explores squeezing-enhanced interferometry in a regime characterized by large phase noise, specifically with fluctuations uniformly distributed in the full $2\pi$ interval.
We consider a differential configuration, where the simultaneous interrogation of multiple atomic sources allows the rejection of arbitrary common-mode noise~\cite{Snadden98, Foster2002, Pereira15}.
Differential schemes are ubiquitous in atom interferometry using uncorrelated atoms~\cite{Rosi15,Barrett2016,Langlois17,Parker18,Rosi19,Hu23,Salvi23,Elliott23}: they have been exploited for precise measurements of fundamental constants~\cite{Fixler07, Lamporesi08, Rosi14}, for demonstrating that gravity induces Aharonov-Bohm phase shifts~\cite{Overstreet22} and for proposing innovative tests of general relativity~\cite{Dimopoulos07, Varoquaux09}, as well as gravitational wave detection in the milli-Hertz range~\cite{Dimopoulos08,Canuel20}. 
Tests of the weak equivalence principle of general relativity have been conducted by measuring the differential free-fall acceleration of distinct atomic species, isotopes, or internal states~\cite{Schlippert14,Zhou15,Asenbaum20,Zhang20,Barrett22,Struckmann24}. 
Differential schemes have also been used for precise measurements of rotations~\cite{Durfee06}, gravity cartography~\cite{Stray2022} and synchronous clock comparison~\cite{YoungNATURE2020, ZhengNATURE2022}.
Finally, it is worth emphasizing that the high sensitivity interferometers that reached SQL sensitivities~\cite{Gauguet09, Sorrentino2014, Janvier22}, as mentioned above, operate in a differential configuration.

We consider a differential interferometer scheme that utilizes combinations of spin-coherent and spin-squeezed states.
For these probe states, the mean value of the measurement outcomes exhibits a sinusoidal dependence on the phase shift.
In the presence of significant large common-mode phase noise, the correlated data from the two interferometers, on average, trace an ellipse. 
Quantum noise, which we compute from first principles, introduces fluctuations around this ellipse.
To estimate the phase shift, we employ a robust method based on ellipse fitting of correlated data~\cite{Foster2002}, a standard technique in differential interferometry that is inherently resistant to additional noise sources. 
This approach extracts the phase shift of interest from the parameters of the ellipse by fitting the data obtained from the two coupled interferometers.
Besides cold atoms, ellipse fitting has been applied to optical Michelson-Morley interferometry~\cite{Collett2014_2,Collett2015} and is implemented across various fields.
Unlike other methods, such as maximum likelihood or Bayesian estimation~\cite{Stockton2007}, this model-free technique does not require a prior calibration of the experimental output probabilities.
Our approach is thus readily experimentally accessible and extends beyond previous theoretical studies, which mainly focused on  spin-squeezed state advantages under vanishing~\cite{EckertPRA2006} or sufficiently small~\cite{Corgier23} phase noise, or addressed the computation of ultimate sensitivity bounds~\cite{Landini14, Gessner2018}.

Incorporating quantum projection noise from entangled states is a central feature of our analysis.
This distinguishes our approach from existing methods~\cite{Collett2014_1, Collett2014_2,Collett2015, Zhu2023, Zhang023, RidleyEPJ2024}, which primarily consider Gaussian fluctuations around a mean ellipse.
Notably, quantum fluctuations contribute to the bias in ellipse fitting methods. 
To address this, we identify a class of spin-squeezed states that simultaneously reduce this bias and enhance the sensitivity in estimating the differential phase shift, outperforming schemes with uncorrelated particles even under significant phase noise.
Specifically, we determine an optimal value of the squeezing strength that leads to nearly unbiased differential phase measurements in a broad range of signal phase shifts $0 \lesssim \delta\phi \lesssim \pi/2$ and provides a sensitivity improvement by a factor $N^{1/6}$ relative to the SQL, where $N$ is the total number of particles.
To emphasize the robustness of the ellipse fitting method, we compare our results with a simplified analytical ellipse model developed in this study, as well as with a fringe fitting method relying on noise correlation with auxiliary sensors, and finally with the optimal phase sensitivity given by the Cram\'er-Rao bound.

\section{Model and Methods}
\label{secII}
 
\subsection{The differential  interferometer scheme} \label{secII.A}

The differential scheme under consideration is shown in Fig.~\ref{fig_1}. 
It consists of two Ramsey interferometers operating in parallel and using a common laser to generate beam splitters and mirrors (not explicitly shown in Fig.~\ref{fig_1}).
Our approach applies to Mach-Zehnder interferometers as well. 
Each interferometer, denoted as $A$ and $B$, is modeled as a two-mode device, $a_1, a_2$ being the modes of interferometer $A$ and $b_1, b_2$ those of interferometer $B$.
The associated bosonic annihilation and creation operators  are respectively $\hat{a}_1, \hat{a}_2, \hat{b}_1, \hat{b}_2$ and $\hat{a}_1^{\dagger}, \hat{a}_2^{\dagger}, \hat{b}_1^{\dagger}, \hat{b}_2^{\dagger}$.
We also introduce collective pseudo-spin operators
$\hat{J}_{M,x} = (\hat{m}_1^\dag \hat{m}_2 + \hat{m}_2^{\dag}\hat{m}_1)/2$, $\hat{J}_{M,y}= (\hat{m}_1^\dag \hat{m}_2 - \hat{m}_2^{\dag}\hat{m}_1)/2i$ and $\hat{J}_{M,z} = (\hat{m}_1^\dag \hat{m}_1 - \hat{m}_2^{\dag}\hat{m}_2)/2$, where $\{m,M\}=\{a,A\}$ or $\{b,B\}$.
These operators satisfy the usual commutation relations $[\hat{J}_{M,i},\hat{J}_{M,j}]=i \epsilon_{ijk} \hat{J}_{M,k}$ with $\epsilon_{ijk}$ being the Levi-Civita symbol.
For simplicity, we assume that both interferometers use a probe state with the same number of particles, $N$. 
The two-mode output state of the interferometer $M$ is~\cite{Yurke86} 
\be 
\label{Eq_psi_out}
\ket{\psi_M^{\rm out}(\phi_M)} = \exp \{-i \phi_{M} \hat{J}_{M,y} \}\ket{\psi_M^{\rm in}},
\ee
where $\phi_M$ is the accumulated phase shift and $\ket{\psi_{M}^{\rm in}}$ is the generic input state.
The transformation $\exp\{-i \phi_{M} \hat{J}_{M,y}\} =\exp\{i (\pi/2) \hat{J}_{M,x}\} \exp\{-i \phi_{M} \hat{J}_{M,z}\} \exp\{-i (\pi/2) \hat{J}_{M,x}\}$ is obtained as a sequence of two beam splitters (collective $x$-rotation) and a phase precession  ($z$-rotation).
In this article, we consider the probe state of each individual interferometer being either a spin-coherent state, described by the binomial particle-number distribution~\cite{Arecchi1972}
\be 
\label{Eq_psi_Coh}
\ket{\psi_M^{\rm Coh}}= \frac{1}{2^{\NT/2}} \sum_{n=0}^{\NT} \binom{\NT}{n}^{1/2}~\ket{\NT-n}_{m_1}\ket{n}_{m_2},
\ee
or a state generated through one-axis twisting (OAT) dynamics~\cite{Kitagawa1993},
\be 
\label{Eq_psi_Sq}
\ket{\psi_M^{\rm Squ}} = \exp\{-i \nu \hat{J}_{M,x}\}\exp\{-i \tau \hat{J}_{M,z}^2\}  \ket{\psi_M^{\rm Coh}}.
\ee
The state Eq.~(\ref{Eq_psi_Sq}) is spin-squeezed~\cite{Wineland1994} for $0< \tau \lesssim 1/\sqrt{N}$, where $\tau$ is the parameter controlling the OAT evolution. 
In Eq.~(\ref{Eq_psi_Sq}), OAT is followed by an appropriate rotation of angle $\nu$ around the $x$ axis to minimize the fluctuations of $\hat{J}_z$.  
In the inset of Fig.~\ref{fig_1}(a) we show the Wigner distribution for different input states considered in this manuscript: the top distribution is for the coherent state of Eq.~(\ref{Eq_psi_Coh}), while the middle and bottom distributions are for Eq.~(\ref{Eq_psi_Sq}) and specific values of $\tau$ (indicated as $\tau_{\rm ref}$ and $\tau^*$, respectively, see below).

In a differential interferometer configuration, 
\begin{subequations}
\begin{align}
   \phi_A &= \phi_{\rm cn}+\delta\phi/2, \label{phiA} \\
   \phi_B &= \phi_{\rm cn} -\delta\phi/2, \label{phiB}
\end{align}
\end{subequations}
where $\phi_{\rm cn}$ is an uncontrolled phase that is common to both interferometers and fluctuates from shot to shot with a probability distribution $\mathcal{P}(\phi_{\rm cn})$.
The differential phase shift $\delta\phi=\phi_A-\phi_B$ is estimated from measurement results of the observable 
\begin{equation}\label{norm_pop_imbalance}
\hat{z}_M=\frac{2\hat{J}_{M,z}}{N},
\end{equation}
corresponding to the relative number of particles between the two output interferometer modes, normalized to the total atom number.
A collection of $\np$ measurements $\{z_{A,j},z_{B,j}\}_{j=1,\dots,\np}$ can be regarded as a set of points in the $z_A-z_B$ plane.
The measurement data are distributed according to 
\begin{equation}
\label{joint_probability}
     \mathcal{P}(z_A,z_B\vert \delta \phi)=\int_0^{2\pi} \ d\phi_{\rm cn} \ \mathcal{P}_0(z_A\vert \phi_A)\mathcal{P}_0(z_B\vert\phi_B) \mathcal{P}(\phi_{\rm cn}),
\end{equation}
where $\mathcal{P}_0(z_M \vert \phi_M)=|\langle z_M|\psi^{\rm out}_M(\phi_M)\rangle|^2$ is the probability to measure the normalized population imbalance $-1\leq z_M \leq 1$ in the output state $|\psi^{\rm out}_M(\phi_M)\rangle$ and $|z_M\rangle$ is the eigenstate of the $\hat{z}_M$ operator with eigenvalue $z_M$.
Using Eqs.~(\ref{phiA}) and (\ref{phiB}), we find that  Eq.~(\ref{joint_probability}) only depends on $\delta \phi$.

The ultimate sensitivity bound of an unbiased estimate of $\delta \phi$ is given by the Cram\`er-Rao bound
\be \label{CRB}
\sigma_{\delta \phi}^{\rm CRB} = (\np F)^{-1/2},
\ee
where $\np$ denotes the number of repeated independent measurements and 
\begin{equation}
\label{eq_Fisher}
    F=\sum_{z_A,z_B}
    \dfrac{1}{\mathcal{P}(z_A,z_B|\delta\phi)}\left(\dfrac{\partial \mathcal{P}(z_A,z_B|\delta\phi)}{\partial \delta\phi}\right)^2
\end{equation}
is the Fisher information computed from the probability distribution Eq~(\ref{joint_probability}), the sums running over $z_{M}=-1, -1+1/N,...,1-1/N,1$.

\subsection{Small-noise environment}\label{secII.B}

In a small-noise environment, where the width of $\mathcal{P}(\phi_{\rm cn})$ is negligible i.e. $\phi_{\rm cn}\ll 2\pi$, we have $\mathcal{P}(z_A,z_B\vert \delta \varphi) \approx \mathcal{P}(z_A\vert \phi_A) \mathcal{P}(z_B\vert \phi_B)$.
The phases $\phi_A$ and $\phi_B$ are estimated independently, such that
\be \label{deltaphiest}
\delta \phi_{\rm est} = \phi_{A, \rm est} -\phi_{B, \rm est},
\ee
where $\phi_{M,{\rm est}}$ is an estimate of $\phi_{M}$.
The corresponding uncertainty, $\sigma_{\delta \phi}^2 = \sigma_{\phi_A}^2+\sigma_{\phi_B}^2$, is given by the sum of estimation variances in each interferometer.
A practical strategy to estimate $\phi = \phi_M$ (we drop the pendix $M$ where unnecessary) is the method of moments, based on inverting the relation
\be \label{z_av}
\bar{z}(\phi)\equiv \langle \psi^{\rm out}(\phi)| \hat{z} |\psi^{\rm out}(\phi)\rangle = - C_{\tau}\sin(\phi), 
\ee
where $C_{\tau}=\cos^{N-1}(\tau)$ is the amplitude of sinusoidal phase oscillations and $\bar{z}$ is estimated by averaging over $\np$ measurements.
The phase is estimated as 
\begin{equation} \label{fitfringe}
    \phi_{\rm est} = \arcsin[-\bar{z}/C_{\tau}].
\end{equation}
The corresponding sensitivity is computed by error propagation, 
\be
\label{errorprop}
\sigma_{\phi}(\phi)= \frac{\sigma_{z}(\phi)}{\sqrt{\np}\vert \partial \bar{z}/\partial \phi\vert},
\ee
where $\sigma^2_{z}(\phi) = \langle \psi^{\rm out}(\phi)| \left[\hat{z} - \bar{z}(\phi)\right]^2 |\psi^{\rm out}(\phi)\rangle$.
We find
\be   \sigma_{z}^2(\phi)=\cos^2(\phi) \ \sigma_{z}^2\vert_{\phi=0}+\sin^2(\phi) \ \sigma_{z}^2\vert_{\phi=\frac{\pi}{2}}, \label{variance_short}
\ee  
with
\begin{subequations} \label{var_phi}
\begin{align}
    & \sigma_{z}^2\vert_{\phi=0}=\frac{1}{N}+\frac{N-1}{4N}\left[\mathcal{K}_1-\sqrt{\mathcal{K}_1^2+\mathcal{K}_2^2}\right],\label{var_phi_0} \\
    & \sigma_{z}^2\vert_{\phi=\frac{\pi}{2}}=\left(1-C_\tau^2\right)-\frac{(N-1)\mathcal{K}_1}{2N},\label{var_phi_pi_2}
\end{align}
\end{subequations}
$\mathcal{K}_1=1-\cos^{N-2}(2\tau)$ and $\mathcal{K}_2=4\sin\tau\cos^{N-2}(\tau)$~\cite{Kitagawa1993}. 
For two coherent states Eq.~(\ref{Eq_psi_Coh}), we have 
\be
\label{eq_SQL_diff}
\sigma_{\delta\phi}^{\rm SQL}  = \sqrt{2} \ \sigma^{\rm SQL}_{\phi} = \sqrt{2}\ \np^{-1/2} N^{-1/2},
\ee
and is indicated as standard quantum limit (SQL) for the differential scheme, where $\sigma^{\rm SQL}_{\phi}$ is the SQL for the single interferometer. 
With squeezed states, it is possible to minimize Eq.~(\ref{errorprop}) with respect to the squeezing strength $\tau$.
The minimum is achieved for $\tau_{\rm ref}\approx 3^{1/6}/N^{2/3}$ in the limit $N\gg 1$ \cite{Pezze2009}. 
From Eq.~\makeref{errorprop}, we have that
\be \label{sigmatauref}
\sigma_{\delta\phi}\vert_{\phi=0,\tau_{\rm ref}}  \approx3^{1/3} \np^{-1/2} N^{-5/6},
\ee
where the optimal phase sensitivity of a single interferometer is obtained at mid-fringe, $\phi=0$ (corresponding to $\bar z =0$).
Figures~\ref{fig_1}(c) and (d) show $\bar z (\phi)$ as a function of $\phi$, for a coherent state, $\tau=0$, and a spin-squeezed state of strength $\tau = \tau_{\rm ref}$ respectively. 
The mean squared error of the population imbalance $\sigma_{z}(\phi)$ is added vertically to $\bar z (\phi)$ and provides the shaded area.
Notice that, according to the error propagation formula, Eq.~(\ref{errorprop}), the phase sensitivity, $\sigma_\phi$ is obtained as the horizontal width of the shaded area.

\subsection{Large noise environment}\label{secII.C}

We refer to large noise environment when $\phi_{\rm cn}$ has a uniform distribution in $[0,2\pi]$, namely $\mathcal{P}(\phi_{\rm cn})=1/(2\pi)$.
In this case, the standard error propagation analysis outlined above cannot be applied since the large phase noise averages out the mean signal of Eq.~(\ref{z_av}).
By exploiting correlations between the output measurements of two interferometers, it is nevertheless possible to estimate the differential phase from the shape of a Lissajous curve.
In fact, when $\phi_{\rm cn}$ spans the full interval $[0, 2\pi]$, and $\bar{z}_M$ has a sinusoidal dependence on the phase as in Eq.~(\ref{z_av}), the average moments $\bar{z}_A$ and $\bar{z}_B$ satisfy the ellipse equation
\begin{equation}
\label{average_ellipse_1}
\dfrac{\bar{z}_A^2}{C_{\tau_A}^2} -2\cos(\delta\phi) \dfrac{\bar{z}_A \bar{z}_B}{C_{\tau_A}C_{\tau_B}}  +\dfrac{\bar{z}_B^2}{C_{\tau_B}^2} -\sin^2(\delta\phi)=0,
\end{equation}
where $C_{\tau_M}$ denotes the fringe contrast for a squeezed state of strength $\tau_M$. 
Equation (\ref{average_ellipse_1}) depends strongly on $\delta\phi$: for $\delta\phi=0$ or $\pi$, Eq.~(\ref{average_ellipse_1}) collapses on a straight line, while for $\delta\phi=\pi/2$ Eq.~(\ref{average_ellipse_1}) is a circle.
Quantum projection noise determines a spreading of measurement data $\{z_{A,j},z_{B,j}\}_{j=1,\dots,\np}$ around the average ellipse.
In particular, the width of the ellipse is dominated by the largest term between $\sigma^2_{z_M}\vert_{\phi=0}$ and $\sigma^2_{z_M}\vert_{\phi=\pi/2}$ in Eq.~(\ref{variance_short}).
Panels (f) and (g) of Fig.~\ref{fig_1} show color plots of the probability distribution, $\mathcal{P}(z_A,z_B\vert\delta\phi)$ of Eq.~\makeref{joint_probability}, for two coherent states, and two spin-squeezed states with $\tau=\tau_{\rm ref}$, respectively.
In particular, when considering 
$\tau_{\rm ref}$, we find that $\sigma^2_{z_M}\vert_{\phi=\pi/2}$ is much larger than $\sigma^2_{z_M}\vert_{\phi=0}$, see Eqs.~(\ref{var_phi_0}) and (\ref{var_phi_pi_2}).
In addition, we emphasize here that for $\tau\gtrsim\tau_{\rm ref}$, the Lissajous curve is no longer an ellipse but rather deforms into a ``butterfly-like shape''.

The dispersion of the data point around the average ellipse is minimized when $\sigma^2_{z_M}\vert_{\phi=0}=\sigma^2_{z_M}\vert_{\phi=\pi/2}$, such that the variance $\sigma^2_{z_M}$ becomes phase-independent. 
This condition is satisfied for a specific squeezing strength $\tau$ given by $\sigma^2_{z_M}\vert_{\phi=0}=\sigma^2_{z_M}\vert_{\phi=\pi/2}$ and leads to~\cite{approximation2}
\begin{equation}
\label{tau_star}
    \tau = \tau^* \approx \left(\dfrac{2}{N^5}\right)^{1/6} \approx \left(\dfrac{2}{3N}\right)^{1/6}\tau_{\rm ref}.
\end{equation}
The corresponding variance is $\sigma^2_{z_M} \approx 2^{-1/3}N^{-4/3}$.
Figure.~\ref{fig_1}(e) shows $z_M$ (line) with mean square error (shaded region) for the case $\tau = \tau^*$.
Using Eq.~(\ref{errorprop}) as in the small-noise case would give the following differential phase sensitivity at mid-fringe:
\begin{equation}
\sigma_{\delta\phi}|_{\phi=0,\tau^*}=\sqrt{2}\sigma_\phi|_{\phi=0,\tau^*}\approx 2^{1/3}~\np^{-1/2} N^{-2/3}.
\label{errorprop_tau*}
\end{equation}
Interestingly, when the fluctuations of $z_M$ are minimized for all $\phi_M$, a gain over the shot noise limit is possible: the price to pay is a less favorable scaling with atom number compared to the optimal squeezing strength, $N^{-2/3}$ rather than $N^{-5/6}$, Eq.~(\ref{sigmatauref}).
As shown in Fig.~\ref{fig_1}(h), $\tau^*$ provides a narrower distribution of data compared to {\it i}) the case of two coherent states ($\tau=0$) and {\it ii}) the case of two optimized spin-squeezed states in a low noise environment ($\tau = \tau_{\rm ref}$).
While the probability density concentrates around the major axis of the ellipse for $\tau=0$ and around the major and minor axes for $\tau=\tau_{\rm ref}$, a trade-off with nearly uniform distribution is obtained for $\tau=\tau^*$.

\subsubsection{Ellipse fitting and statistical analysis} \label{secII.C.1}

A fit of the measurement data aims at recovering the ``true'' ellipse in Eq.~(\ref{average_ellipse_1}) and provides an estimate $\delta\phi_{\rm est}$ of the differential phase.
The figures of merit we consider are both the variance
\be
\sigma^2_{\delta\phi_{\rm est}} = \langle \delta\phi_{\rm est}^2 \rangle - \langle \delta\phi_{\rm est} \rangle^2,
\ee
which quantifies the precision of the estimate, and the deviation of the mean estimate from the true differential phase, 
\be
\mathcal{B}(\delta\phi_{\rm est}) = \langle\delta \phi_{\rm est}\rangle - \delta \phi,
\ee
indicated as bias and quantifying the accuracy.
In the following, we outline two different approaches depending on the {\it a priori} knowledge about the system.

\begin{figure}[t!]
\includegraphics[width=1.0\columnwidth]{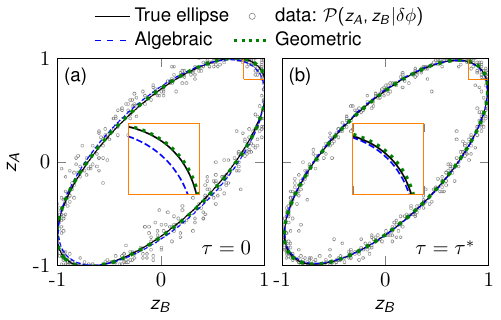}
\caption{Illustration of different ellipse fitting methods for (a) two coherent states and (b) two squeezed states of strength $\tau^*$.
The black line is the ``true'' ellipse given by Eq.~(\ref{average_ellipse_1}). 
The empty gray circles are $\np=500$ randomly data points $\{z_{A,j},z_{B,j}\}_{j=1,\dots,\np}$ sampled according to Eq.~(\ref{joint_probability}).
In each panel, the dashed blue and dotted green line show, respectively, the linear algebraic fit (with trace method) and the non-linear geometric fit.
Here, $\delta\phi=\pi/4$ and $N=100$.
The insets are zooms.
}
\label{fig_2}
\end{figure}

If the squeezing strength is unknown, we cannot make use of Eq.~(\ref{average_ellipse_1}).
We thus fit the data $\{z_{A,j},z_{B,j}\}_{j=1,\dots,\np}$ to the conic function
\be \label{average_ellipse_2}
a\, z_A^2+b\, z_A  z_B+c\, z_B^2+d\, z_A+e\, z_B+f=0,
\ee
to extract the estimate $\delta\phi_{\rm est}$ by means of the relation~\cite{Foster2002},
\begin{equation} \label{theta_ellipse_parameters}
    \delta\phi_{\rm est}=\arccos\left(\frac{-b}{2\sqrt{ac}}\right).
\end{equation}
The  conic parameters $\vect{v} = \{a,b,c,d,e,f\}^\top$ are obtained from a least-square approach that minimizes the sum of squared distances, 
\begin{equation} \label{fit_definition}
    \sum_{j=1}^{\np} d^2(z_{A,j},z_{B,j};\mathcal{C}_{\vect{v}})\new{,}
\end{equation}
between data points and the general conic $\mathcal{C}_{\vect{v}}$, Eq.~(\ref{average_ellipse_2}), with parameter vector $\vect{v}$.
The most popular choices for $d(z_{A,j},z_{B,j};\mathcal{C}_{\vect{v}})$ are the algebraic and the geometric distances~\cite{notefit1}, which give rise to the {\it algebraic} and the {\it geometric} fits~\cite{Gander1994}, respectively. 
The algebraic distance is a linear function of the conic parameter vector $\vect{v}$, whereas the geometric one is non-linear. 
In the algebraic case, it is essential to impose a constraint on the conic parameters. 
In this work, we consider the linear constraints 
$a+c=1$, 
known as the \textit{trace} constraint ~\cite{Rosin1993}, or the quadratic constraint~\cite{Fitzgibbon1996}
$b^2-4ac=-1$,
referred to as \textit{ellipse-specific} constraint~\cite{Gander1980,Halir1998}. 
More details are given in Sec.~\ref{App_B} of the Appendix.
In particular, the algebraic ellipse-fitting method with linear constraints will be indicated as “trace method” in the following.
Other fitting constraints can be found in the literature~\cite{Bookstein1979,Taubin1991,Fitzgibbon1995}.
The estimation depends on the actual form of $d(z_{A,j},z_{B,j};\mathcal{C}_{\vect{v}})$ and on the constraints and, due to the least-square fit on curved lines, it is generally biased~\cite{Kanatani1994,Collett2014_1}.
In Fig.~\ref{fig_2}, we compare the performance of a geometric and an algebraic fit on data points sampled according to Eq.~(\ref{joint_probability}). 
In panel (a), with $\tau=0$, we see that the geometric fit (green dotted line) gets closer to the true ellipse (solid line) than the algebraic one (blue dashed line), at the cost of a non-linear fitting procedure of higher computational complexity.
In contrast, for squeezing strength $\tau = \tau^*$, as shown in panel (b), both fits are close to Eq.~(\ref{average_ellipse_1}), leading to an accurate estimation of $\delta \phi$, as discussed in the following Section.

Instead, if the squeezing strength is known, then we can fit the data to Eq.~(\ref{average_ellipse_1}), where $\delta\phi$ is the only unknown parameter.
We indicate this method as \textit{one-parameter} fit. This provides the estimate:
\be
\delta\phi_{\rm est}=\arccos(h), 
\ee
where $h$ is the real solution to the cubic equation
\begin{equation}\label{cubic}
    G_0 + G_1 h + G_2 h^2 + G_3 h^3 = 0.
\end{equation}
The cubic coefficients $G_0, G_1, G_2$ and $G_3$ are functions of $\{z_{A,j},z_{B,j}\}_{j=1,\dots,\np}$ and require the knowledge of the contrast $C_{\tau_M}$ in Eq.~(\ref{average_ellipse_1}), see Sec.~\ref{App_C} in Appendix.
In the limit of large $\np$, we have
\begin{align}
    & \langle \delta\phi_{\rm est} \rangle \approx f \ (\langle G_0 \rangle, \langle G_1 \rangle, \langle G_2 \rangle, \langle G_3 \rangle),\label{Gau_av_formula}\\
    & \sigma^2_{\delta\phi_{\rm est}} \approx \sum_{j,l=0}^3 \left(\frac{\partial f}{\partial G_j}\right)\textrm{Cov}(G_j,G_l) \left(\frac{\partial f}{\partial G_l}\right).\label{Gau_var_formula}
\end{align}
In the above equations $f(\{ G_i \}_{i\in[0,3]})=\arccos[{f_{cube}\{G_i\}_{i\in[0,3]}}]$ where $f_{cube}(\{G_i\}_{i\in[0,3]})$ corresponds to the Cardano formula expressing the real solution of a cubic equation as a function of its four coefficients $G_0, G_1, G_2, G_3$; $\textrm{Cov}(G_j,G_l)= \langle G_j G_l \rangle -\langle G_j \rangle \langle G_l \rangle$; the partial derivatives on the right-hand side of Eq. \makeref{Gau_var_formula} are evaluated for the arguments $\langle G_0 \rangle, \langle G_1\rangle, \langle G_2 \rangle, \langle G_3 \rangle$ as for the right-hand side of Eq. \makeref{Gau_av_formula}.


\section{Results}
In the following, we show how spin-squeezing enables quantum-enhanced and accurate differential phase measurements in the presence of large common phase noise.
We first present our results focusing on the trace method (namely, algebraic ellipse-fitting method with linear constraints), leading to large bias with coherent states, but numerically very efficient.
We then compare with other methods which are potentially less biased but numerically more demanding, as it is the case of the non-linear geometric fitting method.
We focus on different aspects of quantum-enhanced metrology and analyze the bias, the effective single-point phase sensitivity defined as $\sigma^{\rm eff}_{\delta\phi_{\rm est}}=\sqrt{\np} \ \sigma_{\delta\phi_{\rm est}}$, and the quantum gain with respect to the SQL, defined as $\mathcal{G}=\sigma_{\delta\phi}^{\rm SQL}/\sigma_{\delta\phi}$.
We emphasize the scaling of these quantities with atom number.

\subsection{Bias and phase sensitivity}
\label{secIII.A}

\begin{figure}[t!]
\includegraphics[width=1.0\columnwidth]{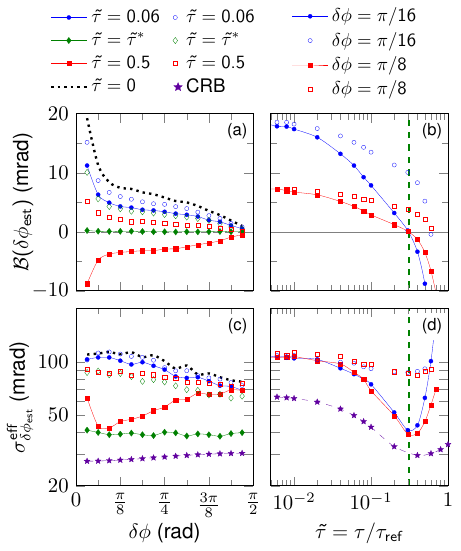}
\caption{
The bias $\mathcal{B}(\delta\phi_{\rm est})$ (top panels) and the effective single-point differential phase sensitivity, $\sigma_{\delta\phi_{\rm est}}^{\rm eff}=\sqrt{\np} \ \sigma_{\delta\phi_{\rm est}}$ (bottom panels) 
are evaluated as a function of the true differential phase $\delta \phi$ (a, c) and as a function of the reduced squeezing strength $\tilde\tau = \tau / \tau_{\rm ref}$ (b, d). 
Different symbols represent calculations for different values of $\tilde\tau$ in (a, c) and $\delta\phi$ in (b, d), as indicated by proper legends. 
In all panels, filled (open) symbols are for the case of two squeezed states (one squeezed and one coherent state) as input, and the dotted black lines in (a,c) hold for two coherent input states, i.e. $\tilde\tau=0$. The filled symbols are joint with the line for clarity purpose. In (c, d), the purple stars show the Cram\`er-Rao bound for the unbiased estimators, Eq.~(\ref{CRB}). In (b, d), the vertical dashed line denotes the optimal squeezing strength $\tau^*$. 
For all data, we employed the algebraic fit method with trace constraint on $\np_{\rm ell}=1000$ ellipses of $\np=1000$ points with $N=500$ atoms per interferometer. Statistical errors are within the size of the symbols.
Note that $\np \gtrsim 10^2$ ensures convergence (see Sec.C of the Appendix).
}
\label{fig_3}
\end{figure}

Previous works~\cite{Foster2002,Pereira15,Rosi15,Barrett2016,Langlois17,Parker18,Rosi19,Hu23,Salvi23,Elliott23} have shown that the bias $\mathcal{B}(\delta\phi_{\rm est})$ in an ellipse fitting procedure in the presence of noise is reduced for $\delta\phi \approx \pi/2$, while being large around $\delta\phi = 0$.
This result is observed for various combinations of probe states, in Fig.~\ref{fig_3}(a).
Using one spin-squeezed state, shown by empty symbols, or two spin-squeezed states, shown by filled symbols, leads to bias reduction for any differential phase when compared to the case of two coherent-states (black dotted line).
In Fig.~\ref{fig_3}(b) we further analyze the bias for two squeezed states of strength $\tau$. 
For all values of the differential phase we analyzed [the cases $\delta \phi = \pi/16$ and $\delta \phi = \pi/8$ are reported in Fig.~\ref{fig_3}(b)], we always found a squeezing strength at which the bias curve crosses the zero-bias line, implying a complete cancellation of the bias. 
Such bias-free configuration depends on $N$ and is realized in the neighborhood of $\tau^*$.
A slight shift of the optimal $\tau$ with respect to $\tau^*$ is a finite-size effect due to the relatively small value of $N$ considered: we have checked that the discrepancy reduces with increasing $N$.
Below, in Sec.~\ref{secIII.B}, we discuss how the bias at $\tau=\tau^*$ scales with $N$.
We anticipate that the bias is much smaller than that found for the coherent state and has a faster decrease with $N$. 
The configuration $\tau=\tau^*$ is shown by the filled green diamonds in Fig.~\ref{fig_3}a and highlighted by the vertical dashed green line in Fig.~\ref{fig_3}b.
An unbiased estimation can be also obtained when only a single spin-squeezed is used but at the expense of larger squeezing strength, $\tau>\tau^*$, as shown Fig.~\ref{fig_3}b.

Figure~\ref{fig_3}(c,d) shows the effective single-point differential phase sensitivity, $\sigma^{\rm eff}_{\delta\phi_{\rm est}}$, following the same color code and symbols as in Fig.~\ref{fig_3}(a,b).
First, the effective single-point differential phase sensitivity improvement with two spin-squeezed states (filled symbols) is significantly larger than that obtained with a single squeezed state (empty symbols) where the gain with respect to the SQL is at best $\sqrt{2}$~\cite{Corgier23}.
Furthermore, the best ellipse fitting result is obtained for two spin-squeezed states of strength $\tau=\tau^*$ where the differential phase sensitivity is essentially independent of the differential phase, as shown by the filled green diamonds in Fig.~\ref{fig_3}c.
Interestingly, in the case $\tau=\tau^*$, high-sensitivity ellipse fitting is possible in the full range $0 \lesssim \delta\phi \lesssim \pi/2$.
The sensitivity of the ellipse fitting for the optimal differential spin-squeezed configuration ($\tau_A = \tau_B = \tau^*$) is found only slightly above the Cram\`er-Rao bound for unbiased estimators, Eq.~(\ref{CRB}), shown as the filled purple stars.
Saturating the Cram\`er-Rao bound would require a more involved parameter estimation analysis, for instance, using a maximum likelihood or a Bayesian method.
However, these approaches require the knowledge of the conditional probability distribution Eq.~(\ref{joint_probability}) for all possible measurement events $z_A$ and $z_B$.
This may be accessed by preliminary calibration of the differential interferometer, which seems rather impractical as it requires collecting a large number of data.

\subsection{Comparison of different ellipse-fitting methods}

\begin{figure}[h!]
\includegraphics[width=1.0\columnwidth]{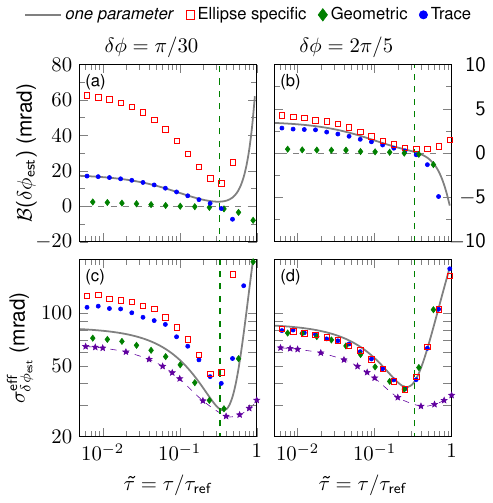}
\caption{Bias, $\mathcal{B}(\delta\phi_{\rm est})$, and effective single-point differential phase sensitivity, $\sigma_{\delta\phi_{\rm est}}^{\rm eff}=\sqrt{\np}\sigma_{\delta\phi_{\rm est}}$, obtained for different ellipse fitting methods:
algebraic with trace constraint used in Fig.~\ref{fig_3} in filled blue circles,
algebraic one-parameter (solid gray line),
algebraic with ellipse-specific constraint in red squares and
geometric in green diamonds.
All methods are discussed in Sec.~\ref{secII.C}.
The vertical dashed line highlights the value of $\tau^*$, see Eq.~(\ref{tau_star}).
The calculations are performed using $\np_{\rm ell} =2000$ ellipses each containing $\np=1000$ points. 
Bias and sensitivity for the one-parameter fit are obtained by evaluation of  Eqs.~\makeref{Gau_av_formula} and \makeref{Gau_var_formula}. 
Here we used $N=500$ atoms per interferometer.
}
\label{fig_4}
\end{figure}

To strengthen the above results, we compare four different ellipse fitting methods in Fig.~\ref{fig_4}.
The \textit{algebraic fit with trace constraint} used in Fig.~\ref{fig_3} is shown as the filled blue circles, the \textit{algebraic fit with ellipse-specific constraint} is shown as the empty red squares, the \textit{geometric fit} is shown as the filled green diamonds and the \textit{one-parameter analytical fit}, is shown by the solid gray line.
For simplicity, we restrict to the case of two spin-squeezed states with identical squeezing strength.

Independently of the fitting procedure, we recover that $\tau\approx\tau^*$ minimizes the bias, as shown in Fig.~\ref{fig_4}(a-b) for two different phase differences, and is the best choice of $\tau$ leading to the optimum phase sensitivity (apart from finite size effects, as discussed above), as shown Fig.~\ref{fig_4}(c-d). 
In particular, only the ``trace'' and ``geometric'' ellipse fitting allow to cancel the bias for any differential phase.
In addition, a direct comparison with the Fisher information, leads to the conclusion that geometric ellipse fitting is the optimal fitting procedure in the regime of small differential phase when $\tau\approx\tau^*$, as shown in Fig.~\ref{fig_4}(c).

\subsection{Bias and phase sensitivity scaling with atom number}
\label{secIII.B}

\begin{figure}[h!]
\includegraphics[width=1.0\columnwidth]{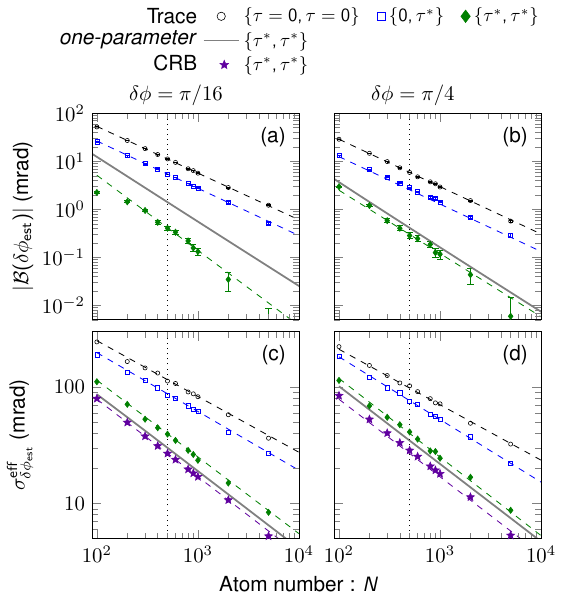}
\caption{Scaling of the bias, $\mathcal{B}(\delta\phi_{\rm est})$ (a-b), and effective single-point differential phase sensitivity, $\sigma^{\rm eff}_{\delta\phi_{\rm est}}=\sqrt{\np} \ \sigma_{\delta\phi_{\rm est}}$ (c-d).
We consider coherent-coherent (black circles), coherent-squeezed (blue squares) and squeezed-squeezed (green diamonds and solid gray lines) input state configurations.
We indicate as $N$ the number of atoms in each interferometer.
Vertical rows correspond to different values of $\delta \phi$.
In panels (a) and (b) the error bars correspond to the statistical error given by $\sigma_{\delta\phi_{\rm est}}/\sqrt{\np_{\rm ell}}$ where $\np_{\rm ell}=1000$ is the total number of ellipses containing each $\np=1000$ points.
The purple stars are Eq.~(\ref{CRB}) for two spin-squeezed states.
In all panels, the squeezing strength is fixed to $\tau^*$ (see Eq.~\ref{tau_star}). 
The vertical dotted lines highlight the configuration with $N=500$ atoms shown Figs.~\ref{fig_3} and~\ref{fig_4}. 
The dashed lines corresponds to a linear fit restricted to the range $300<N<1000$.
The fit coefficients are given in Table.~\ref{tabfit}.
}
\label{fig_5}
\end{figure}

Figure~\ref{fig_5} shows the scaling of the bias absolute value, $|\mathcal{B}(\delta\phi_{\rm est})|$, and of the effective differential phase sensitivity, $\sigma^{\rm eff}_{\delta\phi_{\rm est}}$, as functions of the atom number $N$ in each interferometer.
Here, ellipse fitting is performed using the algebraic fit with trace constraint (empty symbols) and the one-parameter fit (solid lines).
In each panel, empty symbols or solid lines refer to the combinations of: (i) two coherent states (black circles), (ii) one coherent and one squeezed state (blue squares) and (iii) two squeezed states (green diamonds and solid gray lines). 
The squeezing strength is fixed to $\tau^*$.
To account for the statistical errors, the one-sigma error bars on the bias, $\sigma_{\delta\phi_{\rm est}}/\sqrt{\np_{\rm ell}}$, are shown in Fig.~\ref{fig_5}(a-b) for each configuration. Here,
$\np_{\rm ell}$ denotes the total number of sampled ellipses while the number of points per ellipse, $\np$, is already accounted for in the differential phase sensitivity, see Eq.~(\ref{errorprop_tau*}).

In order to derive the scaling laws for the bias and the sensitivity of the fit with trace constraint, we have performed power-law fitting on the relative data. 
For each series of symbols, the dashed line corresponds to a fit of the form, 
\be
\label{eq_fit}
\log_{10}(y) = \alpha - \beta\log_{10}(N),
\ee
where the values of $\alpha$ and $\beta$ are reported in table.~\ref{tabfit}.
Interestingly, the scaling of the bias with atom number is more advantageous for two squeezed states where $\mathcal{B}(\delta\phi_{\rm est}) \propto 1/N^{1.3}$, than for only one or zero squeezed states where $\mathcal{B}(\delta\phi_{\rm est})\propto 1/N$.
In this analysis, the numerical fluctuations caused by the finite number $\np_{\rm ell}$ of sampled ellipses, mostly affecting the points that correspond to the smallest bias values, can hinder a careful evaluation of the scaling exponents. 
Thus, it is worth noting that the same scaling laws are recovered with the one-parameter analytical fit.
Using a first-order Taylor expansion of Eq. \makeref{Gau_av_formula}, the bias can be expressed through the compact formula (see Sec. \ref{App_C} of the Appendix for details),
\begin{equation}
    \mathcal{B}(\delta\phi_{\rm est}) \approx -4\cot{\delta\phi} \ \frac{H_0+H_2 h^2}{1+2h^2}, \label{bias_approximation}
\end{equation}
where $h=\cos(\delta\phi)$ and $H_0$, $H_2$ are functions of $\tau$ and $N$. For $\tau=0$, we have $H_0=-7/(4N)$ and $H_2=1/N$, implying $\mathcal{B}(\delta\phi_{\rm est}) \sim 1/N$ as found above. On the other hand, for $\tau=\tau^*\sim 1/N^{5/6}$, we have $H_0\sim \sigma_z^2 \sim 1/N^{4/3}$ (see also Sec. \ref{secII.C}) and $H_2=0$, thus recovering $\mathcal{B}(\delta\phi_{\rm est}) \sim 1/N^{4/3}$.

\begin{table}[h!]
\centering
\begin{tabular}{|l|l|l|l|l|}
\hline
y & $\delta\phi$ & $\{\tau_A,\tau_B\}$ & $\alpha$ & $\beta$ \\ \hline \hline
\multicolumn{1}{|c|}{\multirow{3}{*}{$|\delta\phi_{\rm est}-\delta\phi|$}} & \multirow{7}{*}{$\pi/16$} & $\{0,0\}$ &  $3.67$ & \multirow{2}{*}{$\approx 1$} \\ \cline{3-4}
\multicolumn{1}{|c|}{} &  & $\{0,\tau^*\}$ & $3.39$  &  \\ \cline{3-5} 
\multicolumn{1}{|c|}{} &  & $\{\tau^*,\tau^*\}$ & $3.84$ & $\approx 1.3$ \\ \cline{1-1} \cline{3-5} 
\multirow{3}{*}{$\sigma_{\delta\phi_{\rm est}}$} &  & $\{0,0\}$ & $3.37$ & \multirow{2}{*}{$\approx 1/2$} \\ \cline{3-4}
 &  & $\{0,\tau^*\}$ & $3.31$ &  \\ \cline{3-5} 
 &  & $\{\tau^*,\tau^*\}$ & $3.38$ & \multirow{2}{*}{$\approx 2/3$} \\ \cline{1-1} \cline{3-4}
$\sigma_F$ &  & $\{\tau^*,\tau^*\}$ & $3.23$ &  \\ \hline
\end{tabular}
\caption{Coefficient of Eq.~(\ref{eq_fit}) for the data shown in Fig.~\ref{fig_5}. 
Notice that scaling coefficients for $\delta\phi=\pi/16$ are identical to those obtained at $\delta\phi=\pi/4$.}
\label{tabfit}
\end{table}

In panel (c) and (d), we recover the scaling of the differential phase sensitivity with atom number discussed above in Sec.~\ref{secII} in the specific case $\phi_M=0$, i.e. at mid-fringe (see Eqs.~\ref{eq_SQL_diff} and \ref{errorprop_tau*}).
In the case where one or zero squeezed states are used, as respectively shown by the blue squares and black circles, $\sigma_{\delta\phi_{\rm est}} \propto N^{-1/2}$, giving a gain $\mathcal{G}=\sigma_{\delta\phi}^{\rm SQL}/\sigma_{\delta\phi}$ that does not scale with $N$.
On the contrary, for two squeezed states we find $\sigma_{\delta\phi_{\rm est}}\propto N^{-2/3}$, and thus $\mathcal{G}\propto N^{1/6}$, as shown by the green diamonds. 
For all configurations, the corresponding phase sensitivity inferred from the Fisher information has  the same scaling, as shown by the purple stars for the specific case of two squeezed states of strength $\tau^*$.
For the latter case, the difference between ellipse and Fisher information is only 1.5\,dB.
We notice that the gain factor $N^{1/6}$ has been also observed previously when studying the performance of a single squeezed state in an atomic clock~\cite{AndrePRL2004}.

\subsection{Ellipse vs Hybridization with classical sensors}

So far, we have studied a differential atom interferometer scheme shown in Fig.~\ref{fig_1} subject to an unknown random common phase noise $\phi_{\rm cn}$. 
Any information about the actual value of $\phi_{\rm cn}$ at each measurement shot was considered unnecessary, as this noise serves to distribute the data points around the ellipse and thus facilitates the conic fitting to estimate the differential phase of interest.
Now we extend the differential scheme of Fig.~\ref{fig_1} by associating it with an additional classical sensor that measures the common phase noise $\phi_{\rm cn}$.  
For instance, the classical sensor can be a seismometer recording the common-mode vibration of the experimental platform that shifts the laser phase with respect to free-falling atoms.
This hybrid quantum-classical configuration was proposed in Ref.~\cite{Pereira15} and demonstrated in Ref.~\cite{Langlois17}.
This method allows to correlate the contribution of the phase noise $\phi_{\rm cn}$ at the $j$-th iteration with both interferometer outputs $z_{M,j}$.
As a consequence, the output data of each single interferometer becomes distributed along sinusoids, similar to the one shown in Fig.~\ref{fig_1}(c-e), with a phase noise limited by the accuracy of the correlation with the classical sensor. 
The inertial phase of interest is then directly extracted by the fringe fit~\cite{Langlois17}, simultaneously providing $\phi_A^{\rm est}$ and $\phi_B^{\rm est}$, and thus $\delta\phi^{\rm est}$ according to Eq.~(\ref{deltaphiest}).

This hybrid method has demonstrated a virtually unbiased extraction of differential phase $\delta\phi$, i.e. $\mathcal{B}(\delta\phi_{\rm est})=0$, with sensitivities close to the SQL when $\delta\phi \simeq 0$ with coherent spin states~\cite{Langlois17}. 
At $\delta\phi=0$, the noise correlation error (stemming, for example, from the finite bandwidth of the auxiliary classical sensor) responsible for the residual differential phase noise $\sigma_{\delta\phi}^{\rm{corr}}\lesssim0.5$~rad barely affects the sensitivity of the fringe fitting method.
Away from this optimal point, however, the sensitivity rapidly degrades with increasing $\sigma_{\delta\phi}^{\rm{corr}}$.
Given the in-applicability of the ellipse fitting approach in the vicinity of $\delta\phi=0$, we emphasize that both methods are complementary since they optimize the sensitivity in different ranges. 

\begin{figure}[t!]
\includegraphics[width=1.0\columnwidth]{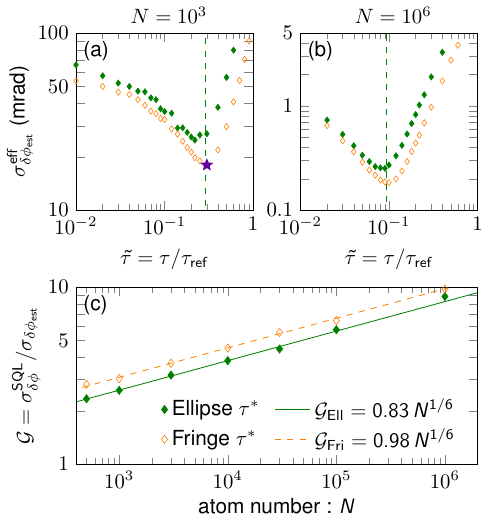}
\caption{
{Ellipse fitting vs fringe fitting.}
In all panels, the configuration \textit{with hybridization} is shown by the empty orange diamonds and obtained by fringe fitting for $\delta\phi=0$.
The configuration \textit{without hybridization} is shown by the filled green diamonds and obtained by ellipse fitting for the specific case $\delta\phi=1$\,rad.
(a-b) Evolution of the effective single shot differential phase sensitivity, $\sigma^{\rm eff}_{\delta\phi_{\rm est}}=\sqrt{\np} \ \sigma_{\delta\phi_{\rm est}}$, for varying squeezing strengths. 
The vertical dashed line denotes the value of $\tau^*$.
The Cram\`er-Rao bound, given by Eq.~\ref{CRB}, is shown as the filled purple stars in (a) for $N=10^3$ atoms. 
(c) Evolution of the quantum gain, $\mathcal{G}=\sigma^{\rm SQL}_{\delta\phi}/\sigma_{\delta\phi_{\rm est}}$, as a function of the atom number, $N$, for $\tau=\tau^*$.
The lines correspond to a fit given in the caption.
The total number of ellipses is $\np_{\rm ell}=1000$ with each $\np=1000$ points.}
\label{fig_6}
\end{figure}

Figure~\ref{fig_6} compares the performance of the ellipse fit method (without hybridization, filled green diamonds) with that achievable using a fringe fit method in the hybrid approach (empty orange diamonds).
The data employed in the fringe fitting method are sampled from the two input squeezed states of $\tau=\tau^{*}$ (optimal for ellipse fitting method) given by Eq.~(\ref{Eq_psi_Sq}), with differential phase $\delta\phi=0$. Here we also restrict the study to the $\sigma_{\delta\phi}^{\rm{corr}}=0$ case, i. e. assuming a perfect knowledge of the $\phi_{\rm cn}$ values.
Note that the differential phase sensitivity with ellipse fitting is independent of the differential phase at $\tau^*$, as shown in Fig.~\ref{fig_3}c and discussed in sec.~\ref{secIII.A}, and we choose $\delta\phi=1$\,rad by default.
The two methods manifest the same scaling with atom number, while the hybridization method provides about 20\% better sensitivity at its optimal point: $\delta\phi=0$.
For a variable squeezing strength $\tau$ in Fig.~\ref{fig_6} (a), the sensitivity values provided by the two methods remain very close up to the optimal squeezing point $\tau^*$, where the fringe fitting sensitivity  reaches the Cram\`er-Rao bound (purple star). 
A somewhat less smooth behavior of the ellipse as compared to fringe fitting results (i.e. larger point-to-point variation) is likely attributed to stronger impact of data dispersion due to small atom number on the ellipse fitting.
The results for a large atom number ($N\gtrsim 10^4$) in Fig.~\ref{fig_6}b are obtained by approximating the distributions $\mathcal{P}_0(z_M\vert \delta \phi_M)$ in Eq.~(\ref{joint_probability}) by Gaussian functions centered in $\bar{z}_M(\phi_M)$, Eq.~(\ref{z_av}), and with variance given by Eq.~(\ref{variance_short}).
Overall, the results confirm the $N^{1/6}$ scaling of the metrological gain, as shown Fig.~\ref{fig_6} (c).

\section{Discussion and Conclusion}

To summarize, we explored the advantages of using squeezed states for estimating differential phase shifts through standard ellipse fitting techniques.
Ellipse fitting offers several key benefits: it requires no calibration of the experimental setup and no prior knowledge of the phase noise model. 
It is inherently robust against large common-phase noise and operates under the minimal assumptions that the mean signal depends sinusoidally on the phase shift and the interferometric setup is affected by large common-mode noise. 
The nonlinear nature of ellipse fitting allows for various techniques, with differing complexities, which we have thoroughly compared in this study.
Our analysis shows the possibility to reach sensitivities below the SQL across a broad signal range $0 \lesssim \delta\phi \lesssim \pi/2$.
Upon optimizing the squeezing strength, we reach a gain in instrument sensitivity over the SQL with an optimal scaling $N^{1/6}$, and also show that spin-squeezing can remarkably suppress the bias in differential phase estimation with respect to using spin-coherent states.
This result is also obtained when considering a hybrid approach where the differential interferometer is assisted by a classical sensor measuring the phase noise value $\phi_{\rm cn}$. 
Our protocol is practical and of immediate applicability in experiments: it uses quantum states that have been demonstrated in several experiments~\cite{Pezze18}. 
The possibility to reach the Heisenberg limit of sensitivity with ellipse fitting may require engineering specific quantum probe states~\cite{KaubrueggerPRX2021,BuzekPRL1999} that, differently from the squeezed states considered in this manuscript, may not be straightforward to generate in atom interferometry experiments.  
As atom interferometers are now reaching control levels dominated by quantum noise~\cite{Gauguet09, Sorrentino2014, Janvier22}, our work represents a significant step toward ultra-precise differential estimation, paving the way for advanced applications in realistic noisy environments. \\

{\bf Acknowledgments.}
This research has been carried out in the frame of the QuantERA project SQUEIS (Squeezing enhanced inertial sensing), funded by the European Union’s Horizon Europe Programme, the Agence Nationale de la Recherche (ANR-22-QUA2-0006).
We also acknowledge financial support from the European Union’s Horizon 2020 research and innovation programme—Qombs Project, FET Flagship on Quantum Technologies Grant No. 820419.
RC, LAS and FP acknowledge the support from a government grant managed by the Agence Nationale de la Recherche under the Plan France 2030 with the reference “ANR-22-PETQ-0005”.
LS and GR acknowledge financial support from the PRIN 2022 project “Quantum Sensing and Precision Measurements with Nonclassical States”.
MM and GMT acknowledge acknowledge funding from the European Union's Next Generation EU Programme IR0000016 I-PHOQS "Integrated Infrastructure Initiative in Photonic and Quantum Sciences.
LS and GMT acknowledge financial support from PNRR MUR Project No. PE0000023-NQSTI

\clearpage

\section{Appendix}
\appendix

\section{Conic fitting}
\label{App_B}

In this section we provide an overview of the ellipse-fitting methods that are relevant to the present work, see Table~\ref{table}.
We will restrict to least-square techniques, defined by Eq. \makeref{fit_definition}.

\begin{table}[!h]
\begin{center}
\begin{tabular}{ |c|c|c| } 
 \hline 
 \bf{Fit Name} & \bf{Description} & \bf{Solution Method} \\
 \hline
Geometric fit & Non-linear & Iterative methods\\ 
& least squares & \\
\hline
 Algebraic fit with & Linear constrained & Solution of an\\ 
 quadratic constraint & least squares &  eigenvalue problem \\
\hline
 Algebraic fit with & Linear constrained & Matrix \\
 linear constraint & least squares & Inversion \\
 \hline
\end{tabular}
\end{center}
\caption{The most popular conic-fitting techniques, all based on a least-squares approach, can be divided into three categories, characterized by different methods: from simple matrix inversion to more involved iterative algorithms with generally slow convergence.
} 
\label{table}
\end{table}
 
\subsection{Geometric fit}

The geometric distance of the point $(z_{A,j},z_{B,j})$ from the conic $\mathcal{C}$ of Eq.~(\ref{average_ellipse_2}) is defined as
\begin{equation}
    d_G(z_{A,j},z_{B,j};\mathcal{C}) = \min_{(z_A,z_B)\in\mathcal{C}}\sqrt{(z_A-z_{A,j})^2+(z_B-z_{B,j})^2}.\label{geometric_distance}
\end{equation}
This optimization problem requires to compute the roots of a fourth-order polynomial \cite{Collett2014_1} which provides a minimum point $(\tilde{z}_{A,j}, \tilde{z}_{B,j}) \in \mathcal{C}$ for each $(z_{A,j},z_{B,j})$ in the data set. 
The fit procedure also requires the minimization of Eq. \makeref{fit_definition} to find the conic $\mathcal{C}$ with optimal parameter vector $\vect{v}=(a,b,c,d,e,f)^T$. 
Overall, the geometric fit involves optimization over a large parameter space, which contains the coordinates of the $\np$ points $(\tilde{z}_{A,j}, \tilde{z}_{B,j})_{j=1,\dots,\np}$ and the parameters in $\vect{v}$. 
No closed-form solution is known, and an iterative optimization procedure is required for convergence. 
Iterative algorithms have been developed to solve general nonlinear least-squares problems, the most famous probably being the Gauss-Newton algorithm and the Levenberg-Marquardt algorithm \cite{Gavin2019}.

\subsection{Algebraic fit}

The algebraic distance of the point $(z_{A,j},z_{B,j})$ from the conic $\mathcal{C}$ of Eq.~(\ref{average_ellipse_2}) is defined as
\begin{equation}
    d_{AL}(z_{A,j},z_{B,j};\mathcal{C}) = \vect{k}_j^T\vect{v},\label{algebraic_distance}
\end{equation}
where $\vect{k}_j=(z_{A,j}^2, z_{A,j}z_{B,j},z_{B,j}^2,z_{A,j},z_{B,j},1)^T$. 
Equation~\makeref{algebraic_distance} is linear with respect to $\vect{v}$, making the algebraic fit a linear least-squares problem. 
In this case, the right-hand side of Eq.~\makeref{fit_definition} can be expressed in matrix form as $\vect{v}^T \vect{S} \vect{v}$: the parameter vector $\vect{v}$ that minimizes this quantity thus determines the fitted conic. 
Here, $\vect{S}=\vect{D}^T\vect{D}$ is a $6\times6$ matrix known as the scatter matrix and $\vect{D}$ is a $m\times6$ matrix known as the design matrix, whose $j$-th row corresponds to the vector of data $\vect{k}_j$ defined above. 
The scatter matrix is symmetric and positive semi-definite:
%
%
a constrained minimization is thus required to avoid the 
trivial solution $\vect{v}=0$. 
%

\subsubsection{Linear constraints}

A linear constraint on $\vect{v}$ can usually be expressed in the form $\vect{w}^T\vect{v}=\alpha$, where $\alpha$ is a scalar and $\vect{w}$ is a 6-dimensional vector. The solution of our least-squares problem is to be found among the stationary points of the Lagrangian function: 
\begin{equation}
    \mathcal{L}(\vect{v},\lambda)=\vect{v}^T \ \vect{S} \ \vect{v}-\lambda(\vect{w}^T\vect{v}-\alpha).
\end{equation}
We obtain the stationary points by solving
\begin{equation}
    \begin{cases}
    &\displaystyle\frac{\partial\mathcal{L}}{\partial \vect{v}}=2\vect{S}\vect{v}-\lambda\vect{w}=0,\\ 
    &\displaystyle\frac{\partial\mathcal{L}}{\partial \lambda}=\vect{w}^T\vect{v}-\alpha=0.\\
    \end{cases}
\end{equation}
Assuming that $\vect{S}$ is invertible (which is true except for sets of points that all lie on the same conic), this admits the unique solution:
\begin{equation}
    \tilde{\vect{v}}=\left(\frac{\alpha}{\vect{w}^T\vect{S}^{-1}\vect{w}}\right)\vect{S}^{-1}\vect{w}.\label{linear_constraint_solution}
\end{equation}
It is possible to see that $\tilde{\vect{v}}$ is not only a stationary point for the Lagrangian function but also minimizes $\vect{v}^T\vect{S}\vect{v}$. 
To show this, we first notice that $\tilde{\vect{v}}^T \vect{S} \tilde{\vect{v}} =\alpha^2/(\vect{w}^T\vect{S}^{-1}\vect{w})$. We then use the Cauchy-Schwartz inequality: $(\vect{w}^T\vect{v})^2\leq(\vect{v}^T \ \vect{S} \ \vect{v})(\vect{w}^T \vect{S}^{-1} \vect{w})$ to obtain $\alpha^2/(\vect{w}^T\vect{S}^{-1}\vect{w})\leq\vect{v}^T \ \vect{S} \vect{v}$ for all $\vect{v}$ such that $\vect{w}^T\vect{v}=\alpha$: this proves our claim.

A common linear constraint is $\vect{w}=(1,0,1,0,0,0)^T$ and $\alpha=1$, giving $a+c=1$, also known as trace constraint. 
%
%
Since this constraint is not ellipse specific, some solutions may not satisfy the condition $b^2-4ac<0$ and must be rejected.

\subsubsection{Quadratic constraints}

A quadratic constraint on $\vect{v}$ can usually be written in the form $\vect{v}^T\vect{W}\vect{v}=\alpha$, where $\alpha$ is a scalar and $\vect{W}$ is a $6\times6$ symmetric matrix. The solution of the least-squares problem is to be found among the stationary points of the Lagrangian function:
\begin{equation}
    \mathcal{L}(\vect{v},\lambda)=\vect{v}^T \ \vect{S} \ \vect{v}-\lambda(\vect{v}^T\vect{W}\vect{v}-\alpha).
\end{equation}
We get the stationary points by solving:
\begin{equation}
    \begin{cases}
    &\displaystyle\frac{\partial\mathcal{L}}{\partial \vect{v}}=\vect{S}\vect{v}-\lambda\vect{W}\vect{v}=0,\\ 
    &\displaystyle\frac{\partial\mathcal{L}}{\partial \lambda}=\vect{v}^T\vect{W}\vect{v}-\alpha=0.\\
    \end{cases}\label{lagrangian_quadratic}
\end{equation}
A necessary condition to solve the constrained minimization problem is that $\vect{v}$ be solution of the generalized eigenvalue problem
\begin{equation} \label{generalized_eigenvalue}
    \vect{S}\vect{v}=\lambda\vect{W}\vect{v}.
\end{equation}
By combining Eqs.~(\ref{lagrangian_quadratic}) and (\ref{generalized_eigenvalue}), we obtain $\vect{v}^T\vect{S}\vect{v}=\lambda\alpha$. 
In addition, from the positivity of $\vect{S}$, we deduce that the only admissible values of $\lambda$ are those with the same sign as $\alpha$; the admissible values for $\alpha$ are instead fixed by the actual form of $\vect{W}$. 

%
%

The ellipse-specific constraint $b^2-4ac=-1$, which we have used in this work, is obtained with $\alpha=1$ and the matrix
\begin{equation}   \vect{W}=\begin{pmatrix}\vect{W}_1&&\vect{0}\\\vect{0}&&\vect{0}\end{pmatrix}, \quad \vect{W}_1=\begin{pmatrix}0&0&2\\0&-1&0\\2&0&0\end{pmatrix},
\end{equation}
where $\vect{0}$ is the $3\times3$ identically vanishing matrix. 
The solution to the ellipse-specific fit is  obtained from 
\begin{equation}
    \tilde{\vect{v}}=\left(\frac{\alpha}{\vect{v}_{\rm max}^T\vect{W}\vect{v}_{\rm max}}\right)^{1/2} \vect{v}_{\rm max},\label{quadratic_constraint}
\end{equation}
where $\vect{v}_{\rm max}$ is any eigenvector for the largest positive eigenvalue of Eq.~\makeref{generalized_eigenvalue} \cite{Gander1980}. 
In this case, Eq.~\makeref{generalized_eigenvalue} admits one positive eigenvalue, and the relative eigenvector $\vect{v}_{\rm max}$ is also unique \cite{Fitzgibbon1996}.
For the numerical implementation of the ellipse-specific fit, we have followed Ref. \cite{Halir1998}: here, by exploiting the block form of $\vect{W}$, Eq.~\makeref{generalized_eigenvalue} was reformulated as a regular eigenvalue problem with a more numerically stable form. 
%

\section{One-parameter fit}
\label{App_C}

The one-parameter fit is of the algebraic type and corresponds to the minimization of Eq. \makeref{fit_definition} with
\begin{equation}
    d(z_{A},z_{B};\mathcal{C}_{\vect{v}}) = C_{\tau_B}^2z_{A}^2+C_{\tau_A}^2z_{B}^2-2C_{\tau_A}C_{\tau_B}z_{A}z_{B} h-C_{\tau_A}^2C_{\tau_B}^2(1-h^2),
\end{equation}
where $d(z_{A},z_{B};\mathcal{C}_{\vect{v}})=0$ corresponds to the average ellipse, $h\equiv \cos(\delta\phi)$ and $\delta\phi$ is the differential phase i.e. the only parameter to be estimated. 
An estimate $\delta\phi_{\mathrm{est}}$ is then obtained by by setting to zero the derivative of Eq. \makeref{fit_definition} with respect to $h$.
This yields the cubic Eq.~\makeref{cubic} with coefficients of the form $G_l=(1/\np)\sum_{j=1}^\np g_l (z_{A,j},z_{B,j})$ where $l=0,1,2,3$ and $g_l$ are functions at most cubic in $z_A$ and $z_B$ with explicit expression:
\begin{align}
    &g_0=\left(C_{\tau_B}^2z_A^2+C_{\tau_A}^2z_B^2-C_{\tau_A}^2C_{\tau_B}^2\right)z_A z_B,\label{g_0_def}\\
    &g_1=-C_{\tau_A}C_{\tau_B}\left(C_{\tau_B}^2z_A^2+C_{\tau_A}^2z_B^2-C_{\tau_A}^2C_{\tau_B}^2+2z_Az_B\right),\label{g_1_def}\\
    &g_2 = 3 \ C_{\tau_A}^2C_{\tau_B}^2 z_A z_B,\label{g_2_def}\\
    &g_3=- C_{\tau_A}^3 C_{\tau_B}^3\label{g_3_def}.
\end{align}
The differential phase estimate is then found as $\delta\phi_{\rm est}=f(G_0,G_1,G_2,G_3)=\arccos[f_{cube}(G_0,G_1,G_2,G_3)]$ where $f_{cube}$ is the real solution of the cubic equation and is given by the Cardano formula.

The statistical analysis of the fit is simplified by recognizing that, since $(z_{A,j},z_{B,j})$ are independent samples with the same distribution $\mathcal{P}(z_A,z_B|\delta\phi)$, then the $g_l (z_{A,j},z_{B,j})$ are just independent samples of $g_l (z_A, z_B)$ with identical distributions.
Additionally, the estimator for $\delta\phi_{\rm est}$ is a function of sample moments which implies that the approximate Eqs.~\makeref{Gau_av_formula} and \makeref{Gau_var_formula} for the mean value and the variance, respectively, are valid in the limit of large $\np$ \cite{Hurt1976}. 
We use the identities $\langle G_l \rangle=\langle g_l \rangle$ and $\textrm{Cov}(G_j,G_l)=(1/\np) \ \textrm{Cov}(g_j,g_l)$ to further reduce the statistical analysis to the calculation of the first and second moments of $g_l$, which are independent of $\mathcal{N}$. The mean values $\langle g_l \rangle$ are given by the combination of the average over the common phase noise $\phi_{\rm cn}$ and of the quantum-mechanical expectation value:
\begin{equation}
    \langle g_l \rangle=\frac{1}{2\pi}\int_0^{2\pi} d\phi_{\rm cn}\langle\psi^{\rm out}_A(\phi_A)|\langle\psi^{\rm out}_B(\phi_B)|\hat{g}_l|\psi^{\rm out}_A(\phi_A)\rangle|\psi^{\rm out}_B(\phi_B)\rangle,\label{integral}
\end{equation}
where $\phi_A=\phi_{\rm cn}+\delta\phi/2$ and $\phi_B=\phi_{\rm cn}-\delta\phi/2$. Here, $\hat{g}_l$ is the operator equivalent of $g_l$ given by $\hat{g}_l=g_l(\hat{z}_A,\hat{z}_B)$ which does not require symmetrization since $\hat{z}_A$ and $\hat{z}_B$ commute. Analogous formulas can be derived for the second moments.

\subsubsection{Squeezed-spin-state moments}

The first moment of $\hat{g}_l$ requires the evaluation of moments of $\hat{z}$ up to third order and the second moments involve moments of $\hat{z}$ up to the sixth order. For the calculation we refer to a single squeezed state, thus suppressing subscripts $A$ and $B$, and to the angular momentum operator $\hat{J}_z = N\hat{z}/2$.

Using the Heisenberg picture, we write $\langle \psi^{\rm out}(\phi)|\hat{J}_z^k|\psi^{\rm out}(\phi)\rangle=\langle \psi^{\rm Coh}|\hat{J}_z^k(\phi,\nu,\tau)|\psi^{\rm Coh}\rangle$, where $|\psi^{\rm Coh}\rangle$ is the coherent spin state of Eq. \makeref{Eq_psi_Coh} and $\hat{J}_z(\phi,\nu,\tau)$ is the transformed $\hat{J}_z$ for a rotated squeezed spin state, according to the unitary $\textrm{exp}(-i\phi\hat{J}_y) \ \textrm{exp}(-i\nu\hat{J}_x) \ \textrm{exp}(-i\tau\hat{J}_z^2)$. Simple formulas for SU(2) rotations and one-axis twisting \cite{Kitagawa1993,PuriBOOK} allow us to find:
\begin{equation}
    \hat{J}_z(\phi,\nu,\tau)=\cos\phi  \left[\cos\nu\hat{J}_z+\sin\nu\hat{J}_y(\tau)\right]-\sin\phi \ \hat{J}_x(\tau),\label{Jz_trasformation}
\end{equation}
where $\hat{J}_{x}(\tau)=[\hat{J}_+ \ \textrm{exp}_+(\tau)+\textrm{exp}_-(\tau) \ \hat{J}_-]/2$, $\hat{J}_{y}(\tau)=[\hat{J}_+ \ \textrm{exp}_+(\tau)-\textrm{exp}_-(\tau) \ \hat{J}_-]/(2i)$, $\hat{J}_{\pm}=\hat{J}_x \pm i\hat{J}_y$ and $\textrm{exp}_{\pm}(\tau)=\textrm{exp}[\pm 2i\tau(\hat{J}_z+1/2)]$. With these relations, $\hat{J}_z(\phi,\nu,\tau)$ can be written in terms of $\hat{J}_z,\hat{J}_+,\hat{J}_-$ and $\exp_{\pm}(\tau)$ only. Any expectation value on $|\psi^{\rm Coh}\rangle$ that contains the transformed $\hat{J}_z$ operator can then be computed using the derivatives with respect to their arguments of the normally-ordered generating function \cite{Arecchi1972}
\begin{align}
    X_N(\alpha,\beta,\gamma)&=\langle\psi^{\rm Coh}|\exp\left(\alpha\hat{J}_+\right)\exp\left(\beta\hat{J}_z\right)\exp\left(\gamma\hat{J}_-\right)|\psi^{\rm Coh}\rangle \nonumber\\
    &=\left(\frac{1}{2}\right)^N\left[e^{\beta/2}+e^{-\beta/2}(\alpha+1)(\gamma+1)\right]^N
\end{align}
and of the anti-normally-ordered $X_A(\alpha,\beta,\gamma)$ similarly defined. In the above equation, $N$ is the number of atoms. 
The expectation value of Eq. \makeref{Jz_trasformation} on $|\psi^{\rm Coh}\rangle$ is the sum of two terms: the one proportional to $\cos\phi$ is found to vanish, and $\langle \psi^{\rm Coh}|\hat{J}_x(\tau)|\psi^{\rm Coh}\rangle=N\cos^{N-1}(\tau)/2$ for the other term. By rescaling by $2/N$, we recover Eq. \makeref{z_av}. 
By squaring Eq. \makeref{Jz_trasformation}, one will obtain three terms: one proportional to $\cos^2\phi$, another to $\sin^2\phi$, and, finally, one proportional to  $\sin\phi\cos\phi$, whose expectation value on $|\psi^{\rm Coh}\rangle$ is also found to vanish. 
We thus recover the functional dependence of Eq. \makeref{variance_short}, obtained from the variance of $\hat{J}_z$, on the phase shift $\phi$. The two oscillation extrema $\sigma^2_{\hat{z}}|_{\phi=0}$ and $\sigma^2_{\hat{z}}|_{\phi=\pi/2}$ correspond to the variance on $|\psi^{\rm Coh}\rangle$ of $\cos\nu\hat{J}_z+\sin\nu\hat{J}_y(\tau)$ and of $\hat{J}_x(\tau)$, respectively, properly rescaled by $4/N^2$.

\subsubsection{Fit bias}

The method outlined above also allows to find expressions for $\langle g_l\rangle$ which can be used for the analytical study of the fit bias $\mathcal{B}(\delta\phi_{\rm est})$. The calculation further requires the evaluation of the integral in Eq. \makeref{integral}. 
Assuming $\tau_A=\tau_B\equiv\tau$, we obtain:
\begin{align}
\langle g_2 \rangle=& \frac{3}{2} C_{\tau}^6 h,\\
\langle g_1 \rangle=&-\left\{\frac{C_{\tau}^4}{4}+\frac{C_{\tau}^2}{2}\left(5\sigma_0^2+3\sigma^2_{\pi/2}\right)+\frac{\left(\sigma_0^2-\sigma_{\pi/2}^2\right)^2}{4}-2\sigma_0^2\sigma_{\pi/2}^2\right.\nonumber\\
&+\left.\left[\frac{C_{\tau}^4}{2}-C_{\tau}^2\left(\sigma_0^2-\sigma_{\pi/2}^2\right)+\frac{\left(\sigma_0^2-\sigma_{\pi/2}^2\right)^2}{2}\right]h^2\right\}C_{\tau}^2,\\
\langle g_0 \rangle=&C_{\tau}^3\left[-\frac{C_{\tau}^3}{2}+\frac{3}{4N}C_{\tau}+\frac{3\Sigma_1+\Sigma_2}{4}\right]h.
\end{align}
In these equations, $\sigma_0^2$ and $\sigma_{\pi/2}^2$ are shorthand notation for $\sigma_z^2|_{\phi=0}$ and $\sigma_z^2|_{\phi=\pi/2}$, and, in the limit $N\gg1$, $\Sigma_1=(C_{3\tau}+3C_{\tau})/4+3C_{\tau}/N$ and $\Sigma_2=-3[c_{\nu}^2 \sin^2(\tau) \ C_{\tau}+s_{\nu}^2(C_{3\tau}-C_{\tau})/4-s_{\nu}c_{\nu}\sin(2\tau) \ C_{2\tau}]$ with $c_{\nu}\equiv\cos\nu$, $s_{\nu}\equiv\sin\nu$.

In order to obtain a compact expression for the bias, we perform a Taylor expansion of $f(\braket{g_0},\braket{g_1},\braket{g_2},\braket{g_3})$ about the unbiased point corresponding to $\tau=0$ and $N\rightarrow \infty$. This situation is described by $ g_{\infty} = (\langle g_0 \rangle_{\infty}, \langle g_1 \rangle_{\infty}, \langle g_2 \rangle_{\infty}, \langle g_3\rangle_{\infty})$, where $\langle g_l \rangle_{\infty} = \lim_{N\to \infty} \langle g_l \rangle|_{\tau=0}$. We get $\langle g_0 \rangle_{\infty}=h/4$, $\langle g_1\rangle_{\infty}=-(1/4+h^2/2)$, $\langle g_2 \rangle_{\infty}=3h/2$ and $\langle g_3 \rangle_{\infty}=-1$. The solution of the cubic equation is then, as expected, the unbiased phase difference, i.e. $f(g_{\infty})=\delta\phi$. A first order Taylor expansion of the fit bias centered at $g_{\infty}$ then yields $\mathcal{B}(\delta\phi_{\rm est}) = \langle \delta\phi_{\rm est} \rangle -\delta\phi \approx \sum_{l=0}^3 (\partial f/\partial g_l)_{g=g_{\infty}}(\langle g_l \rangle - \langle g_l \rangle_{\infty})$. The derivatives of $f$ are given by $\partial f/\partial g_l= -[1-(f_{cube})^2]^{-1/2} (\partial f_{cube}/\partial g_l)$ and the derivative of $f_{cube}$ can be evaluated by the implicit function theorem, which gives $(\partial f/\partial g_l)_{g=g_{\infty}}=-4(1-h^2)^{-1/2}(1+2h^2)^{-1} h^l$. We also calculate the difference $\langle g_l \rangle - \langle g_l \rangle_{\infty}$, and, bringing all results together,  obtain Eq. \makeref{bias_approximation}.

\begin{figure}[b!]
\includegraphics[width=0.5\textwidth]{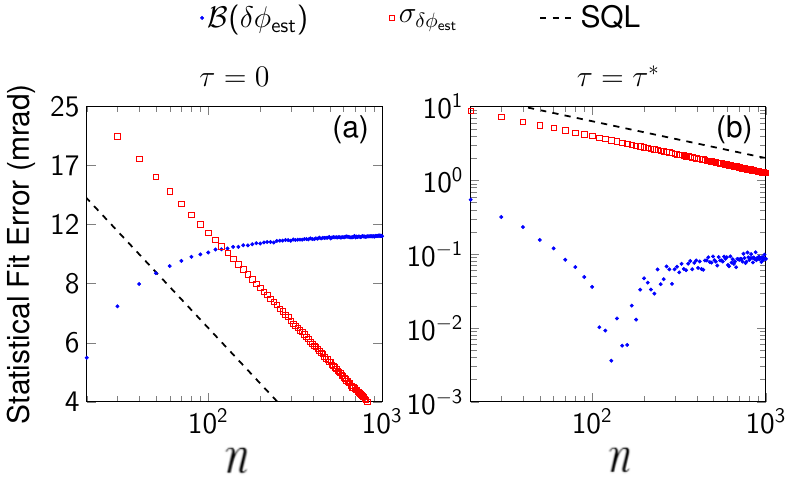}
\caption{Bias $\mathcal{B}(\delta\phi_{\rm est})$ and standard deviation $\sigma_{\delta\phi_{\rm est}}$ as functions of $\np$. Results are relative to $N=500$ atoms, a differential phase shift $\delta\phi=\pi/16$ and are obtained by means of an algebraic fit with trace constraint. The dashed line corresponds to the SQL for the considered differential scheme: $\sqrt{2/(\np N)}$.}
\label{fig_7}
\end{figure}

\section{Dependence on the number of fitted points}

Here, we study the dependence of the fit parameters on the number $\mathcal{N}$ of points in a single elliptical sample, see Fig.~\ref{fig_7}.
There, we plot $\mathcal{B}(\delta\phi_{\rm est})$ (blue points) and $\sigma_{\delta\phi_{\rm est}}$ (red squares) as functions of $\np$, with $\tau=0$ in (a) and $\tau=\tau^*$ in (b). 
While the standard deviation shows the expected scaling $\sim 1/\sqrt{\np}$, the bias saturates for sufficiently large $\mathcal{N}$ to a value dependent on $\delta\phi, \tau$ and $N$ (as studied in the main text). 
While for $\tau=0$ the bias shows limited variation with $\np$ (well within one order of magnitude), for $\tau=\tau^*$ it has a minimum for $\np \approx 10^2$, dipping to almost two order of magnitudes below the saturation value for $\np=10^3$. 
We also notice that the favorable situation where the overall fit error is dominated by the standard deviation --- so that the method can be considered practically unbiased --- can be realized with squeezed states on a wider range of $\np$ values. 
A quantitative formulation of this result in the limit $\np \to \infty$ can be obtained by making use of the scaling laws determined in Sec. \ref{secIII.B}. 
The condition $\mathcal{B}(\delta\phi_{\rm est})<\sigma_{\delta\phi_{\rm est}}$ (where both quantities are now intended in the limit $\np \to \infty$) translates to (i) $\np < N$ in the case of coherent states, as $\mathcal{B}(\delta\phi_{\rm est})\sim 1/N$ and $\sigma_{\delta\phi_{\rm est}}\sim 1/\sqrt{\np N}$, and to (ii) $\np <N^{7/3}$ for squeezed states with $\tau=\tau^*$, as $\mathcal{B}(\delta\phi_{\rm est})\sim 1/N^{4/3}$ and $\sigma_{\delta\phi_{\rm est}}\sim 1/(\sqrt{\np} N^{1/6})$. 
The validity of the upper bounds just derived for $\np$ is confirmed by Fig. \ref{fig_7}: the values $\np=10^3$ and $N=500$ violates $\np<N$ in panel (a) and fulfill $\np<N^{7/3}$ in panel (b).


\begin{thebibliography}{99}
\bibitem{approximation2}
An exact analytical solution to $\sigma^2_{z}\vert_{\phi=0}=\sigma^2_{z}\vert_{\phi=\pi/2}$ can be obtained in the regime $1/N\ll\tau\ll1/\sqrt{N}$, with $N\gg1$. 
%
We use $\sigma^2_{z}\vert_{\phi=0} \approx 1/(2 N^2\tau^2)+N^2\tau^4/12$ and $\sigma^2_{z}\vert_{\phi=\pi/2} \approx N^3\tau^4/4$, giving Eq.~\makeref{tau_star}.
\end{thebibliography}
\end{document}